\documentstyle [12pt,epsfig]{article} 
\makeatletter

\@addtoreset{equation}{section}
\makeatother
\renewcommand{\theequation}{\thesection.\arabic{equation}}
\newcommand{\ba}{\begin{eqnarray}}
\newcommand{\ea}{\end{eqnarray}}

\newcommand{\Rc}{{\cal R}}
\newcommand{\Dc}{{\cal D}}
\newcommand{\AD}{A_{{\rm det}}^{\gamma}}
\newcommand{\ADE}{A_{{\rm det}}^{{\rm e}}}
\begin{document}
\newcommand{\BS}{\bigskip}
\newcommand{\SECTION}[1]{\BS{\large\section{\bf #1}}}
\newcommand{\SUBSECTION}[1]{\BS{\large\subsection{\bf #1}}}
\newcommand{\SUBSUBSECTION}[1]{\BS{\large\subsubsection{\bf #1}}}

\begin{titlepage}
%\hspace*{8cm} {UGVA-DPNC 1998/04-176 April 1998}
%\hspace*{8cm} {arXiv:physics/0606188 v1, June 2006}
\begin{center}
\vspace*{2cm}
{\large \bf Description of diffraction grating experiments for photons
  and electrons in Feynman's space-time formulation of quantum mechanics:
  The quantum origins of classical wave theories of light and massive
  particles}  
\vspace*{1.5cm}
\end{center}
\begin{center}
{\bf J.H.Field }
\end{center}
\begin{center}
{ 
D\'{e}partement de Physique Nucl\'{e}aire et Corpusculaire
 Universit\'{e} de Gen\`{e}ve . 24, quai Ernest-Ansermet
 CH-1211 Gen\`{e}ve 4.
}
\newline
\newline
   E-mail: john.field@cern.ch
\end{center}
\vspace*{2cm}
\begin{abstract}
    The five laws of relativistic quantum mechanics, according to Feynman's path integral formulation,
    are concisely stated and applied to experiments. Reflection diffraction grating
    experiments for both photons and electrons are analysed, in particular the Davisson-Germer experiment
    in which the wave-like property of electrons was first established. It is shown how classical, purely
    spatial, effective wave theories for both photons and electrons are predicted by the path integral formulation
    of quantum mechanics. The standard Copenhagen interpretation of wave mechanics is critically discussed in the
    light of the described experimental applications of the path integral formulation.


\vspace*{1cm}
%\par Published in: {\it American Journal of Physics {\bf 68} (2000), 267-274.} 
\end{abstract}
%\vspace*{4cm}cp
%24 pages, 14 figures, 7 tables
\end{titlepage}  
 
\SECTION{\bf{Introduction}}

         Feynman's path integral formulation of quantum mechanics~\cite{FT,FeynRMP,FH} based on cited earlier
             work of Heisenberg~\cite{Heis1} and Dirac~\cite{DiracLQM} has two distinct parts. The first is a set of
              rules (the laws I--V of Section 2 below) concerning the construction  and interpretation of 
            probability amplitudes for space-time experiments in quantum mechanics\cite{FeynRMP}.
        These rules are valid in
          both the non-relativistic and relativistic theories. The second part is the detailed mathematical development
           of the non-relativistic limit~\cite{FH}.
           The most detailed working-out of Feynman's space-time concepts for quantum mechanics is to be found,
            not in the research literature or text books, but in the popular book `QED the strange story of light
        and matter'
            ~\cite{FeynQED1} published shortly before his death. The photons of which light consists are, of course,
              unlike the particles considered in Refs.~\cite{FT,FeynRMP,FH}, ultra-relativistic.
        In this book, many experiments on propagation, reflection, refraction,
           diffraction and interference of light, which are conventionally described by the classical wave theory,
            are all treated as applications of Feynman's path integral formulation of quantum mechanics. Although 
            a complete physical description of the experiments was given (neglecting only polarisation
             effects) no equations were employed. Many of the experiments described in this book have been worked out
            in full mathematical detail in a previous paper~\cite{JHFAP} by the present author.
            The present paper also describes physical optics experiments for both photons
           and massive particles (electrons) in a similar manner to Ref.~\cite{JHFAP}, but has more avowedly
           pedagogical goals. In particular, the path integral analysis of the experiments, is confronted, in a 
           critical manner, with typical concepts of the `Copenhagen Interpretation' that are found in quantum
            mechanical text books.
             \par Before presenting a space-time analysis\footnote{As in Refs.~\cite{FeynQED1,JHFAP} polarisation
              effects are neglected throughout
             the present paper.} of the original Davisson-Germer experiment~\cite{DG}
                in which `matter waves' were discovered, a reflection diffraction grating experiment with
                 a similar geometry, but using photons produced in the decays of excited atoms, is analysed.
                Although the two experiments are explained similarly by classical, spatial, wave theories
                with similarly defined phenomenological de Broglie wavelengths, the underlying space-time
               physics is seen to be entirely different in the two cases. This distinction was previously
             pointed out in Ref.~\cite{JHFAP} for the case of Young double slit experiments using
             photons or electrons. 
            \par The plan of this paper is as follows: In the following section Feynman's rules for constructing
              and interpreting probability amplitudes~\cite{FT,FeynRMP,FH,LLB,JHFAP} are reviewed. In Section 3
               the space-time propagator for a free relativistic particle is derived. The following three sections
                present space-time analyses of different experiments: in Section 4, one in which a photon, produced
              in the decay of an excited atom, is detected, in Section 5, an experiment with a similarly produced photon
              scattered from a reflection diffraction grating, and in Section 6 a similar experiment using
              an electron beam ---the Davisson-Germer experiment. In Section 7 it is shown,
               following, Ref.~\cite{JHFAP}, how the path integral formulation of quantum mechanics leads to 
                similar, purely spatial, classical wave theories for both photons and massive particles,
               in spite of completely different underlying space-time processes in the two cases.
                The final section contains a critical discussion, in the light of
                  the experiments presented,  of some basic concepts within the standard
               `Copenhagen Interpretation'
                ~\cite{BohrCI,HeisCI} of wave-mechanics, such as wave packets and uncertainty relations.
                 Finally the interpretation of the famous `Schr\"{o}dinger's cat' experiment
               ~\cite{Schrodcat} within the path integral formulation is considered.

  \SECTION{\bf{Feynman's conceptual formulation of quantum mechanics}}
   Feynman's path integral formulation of quantum mechanics is the development of earlier
    work by Heisenberg and Dirac in which the axiomatic basis of the formulation was already
    specified. Heisenberg's work~\cite{Heis1}, concerned the manner in which probability
    amplitudes are to be constructed and interpreted. Dirac's seminal paper `The Lagrangian
     in Quantum Mechanics'~\cite{DiracLQM} provided the dynamical foundation of the theory by specifiying
   the connection between a quantum mechanical path amplitude and the Lagrangian function of
    the corresponding classical system. This work of Heisenberg and Dirac is valid in both
   the non-relativistic and relativistic limits of the theory. Feynman chose, in his original
   path integral paper~\cite{FeynRMP}, to consider only the non-relativistic case for detailed
   mathematical treatment. However, Section 2 of this paper, `The superposition of probability
   amplitudes', reviewing the earlier work of Heisenberg~\cite{Heis1}, is applicable also to
   the relativistic theory. The construction of probability amplitudes for a space-time 
   quantum mechanical experiment, and how such an experiment differs from the corresponding classical
   experiment, which, according to Heisenberg, `when stated in
   a sufficiently general form' is `the center of the whole quantum theory'~\cite{Heis1},
   will now be described.   
   \par Suppose that some quantum mechanical system is prepared in the state $|i\rangle$ and measured
   to be in the state  $|f\rangle$, having, at some intermediate time, passed through the state $|k\rangle$
    \footnote{For concretness, following Feynman~\cite{FeynRMP}, the case of a particle initially at
     some spatial position in the state $|x_i\rangle$ and detected 
     in another $|x_f\rangle$  having passed through an intermediate state  $|x_k\rangle$ may be considered,
     though the formula (2.1) is of quite general validity.}. The probability amplitude
     to measure the state $f$ given the prepared state $i$ is:
     \begin{equation}
      A_{fi}^k \equiv \langle f|k|i \rangle = \langle f|k\rangle\langle k| i\rangle
       \equiv A_{fk}A_{ki}
     \end{equation}
     where the Dirac Bra-Ket notation~\cite{DiracBK} for states and transition amplitudes is employed.
      In this formula the amplitudes $A$ are complex numbers determined by the underlying 
     physics of the process considered. They are typically either space-time propagators
     giving the amplitude to find a particle at some position when it is at a known position
     at some earlier time, or the amplitude for a  particle scattering or production process.
     \par According to the Born rule~\cite{Born} the probability to observe the state $f$ given the 
      states $i$ and $k$ is:
     \begin{equation}
      P_{fi}^k = |\langle f|k|i \rangle|^2 =  |\langle f|k\rangle|^2|\langle k| i\rangle|^2 
     =   P_{fk} P_{ki}. 
     \end{equation}
       This formula expresses the law of conditional probability: the probability of $f$ 
     given $k$ and $i$ is the probability of $k$ given $i$ times the probability of $f$ 
    given $k$. If $k$ takes, with equal probabilities, several different, but unknown, values
    the overall probability $P_{fi}^{\rm{Cl}}$ (where the suffix $\rm{Cl}$ stands for `classical')
     of $f$, given $i$, is:
     \begin{equation}
     P_{fi}^{\rm{Cl}} = \sum_k  P_{fi}^k = \sum_k  P_{fk} P_{ki} =  \sum_k |A_{fk}|^2|A_{ki}|^2.
     \end{equation}
     i.e. the overall probability is the sum of the conditional probabilities. In quantum mechanics,
    the formula (2.3) for the case that the intermediate state  $|k\rangle$ is one of a set specified
    by the label $k$, but the value of $k$ is unknown, is replaced by:
     \begin{equation}
     P_{fi}  = | A_{fi}|^2 =  |\sum_k A_{fi}^k|^2 =  |\sum_k  A_{fk}A_{ki}|^2
     \end{equation}
      where 
   \begin{equation}
         A_{fi} \equiv \sum_k A_{fi}^k = \sum_k A_{fk} A_{ki}
      \end{equation}
    expresses the {\it principle of superposition of the path amplitudes $A_{fi}^k$}.  
    \par The difference between (2.3) and (2.4) and the non-intuitive nature of the latter 
       equation constitute what Feynman called (as exemplified in the Young double slit 
       experiment) the {\it only} mystery of interpretation of quantum mechanics~\cite{FeynYDS}.
    It is the same as what Heisenberg earlier called `the center of the whole quantum theory'~\cite{Heis1}.
     \par Although it is not possible to `explain' the fundamental formula (2.4) in terms
      of our understanding of space, time and causality  in the world of everyday experience,
      it is at least possible to {\it state} how the classical formula (2.3) differs from the
      quantum mechanical one (2.4) that describes the real world. 
      Both (2.3) and (2.4) make predictions for probabilities concerning a statistical
      ensemble of experiments in which $k$ is allowed to vary for fixed values of $i$ and $f$.
      Both predictions are `probabilistic'. It is not the
      case that the classical formula is deterministic and the quantum one probabilisitic.
      However, the classical formula is probabilistic because the value of $k$, although existing,
      (i.e. corresponding to a physically existing trajectory of a particle or a definite time-ordered
      sequence of space-time events) is not known, whereas for the quantum formula
      the value of $k$ is not simply unknown,
      {\it but in principle unknowable}, without destroying the assumed experimental conditions.
      The probabilistic nature of the quantum formula is therefore not a consequence of simple
      ignorance of the values of some actually-existing physical parameter. 
      The logical basis of (2.3) is that, in the different experiments of the ensemble, 
       each one corresponds to a unique value of $k$. Say $k = k'$ for one experiment
         and $k = k''$ for another, corresponding, according to (2.2) to the
        conditional probabilities $P_{fi}^{k'}$ and $P_{fi}^{k''}$, which are added in as 
       contributions to the overall probability in (2.3). That is, each value of $k$
      corresponds to a distinct causal chain in space-time which may be called a `classical
      history'. On the other hand, in (2.4), different values of $k$  {\it contribute to every 
      measurement of} $f$, as if, loosely speaking, the quantum system is simultaneously 
       occupying all possible intermediate states allowed by the experimental configuration.
      However, the conventional notions of space and
       time must still be applied to calculate correctly the path amplitudes $A_{fi}^k$, i.e. 
       particles of known identity and kinematical properties are assumed to be produced, destroyed
       or scattered and {\it to propagate in space-time according to the classical laws of space-time
       geometry and kinematics\footnote{`classical' is used here in the sense of `non-quantum'.
         The kinematical formulas are, in general, those of relativitisic, not classical, mechanics.}}.
    Particle concepts are therefore essential to calculate the path amplitudes
       and any `wave' concept is irrelevant. This will become clear in the specific space-time
       experiments to be discussed in the following sections. When the path amplitudes 
       $A_{fi}^k$ are combined, according to the superposition law (2.5), to give the quantum
      probability amplitude $A_{fi}$, the modulus squared of which gives the probability to
       measure the state $|f \rangle$ when the state $|i \rangle$ is prepared, but the experimental
       configration allows all members of the set of intermediate states  $|k \rangle$,
       there is a breakdown of the correspondence between
        different values of $k$ and different classical histories which occurs in (2.3).
        Quantum mechanical superposition requires that amplitudes (not probabilities) for different
       classical histories are added to form the overall probability amplitude for the experiment,
        like the simultaneous overlapping of different melodies (each one musically coherent) 
        in musical counterpoint. It is as though the unique causal chain of classical physics is replaced,
         in constructing the probability amplitude, by an, in general, infinite number of such
         chains in one-to-one correspondence with the path amplitudes when a {\it single measurement}
        of the final state $|f \rangle$ is performed. This is reminiscent of
         Everett's `Many Worlds' interpretation~\cite{Everett} except that the correspondence is 
        between the one unique actual world described by quantum mechanics and all the different
        distinct classical histories that must be considered in the calculation of the 
        overall probability amplitude for an experiment. The essentially non-classical (and counter intuitive)
        aspect of the situation is that a correspondence exists between a {\it single} observed 
        quantum system (for example the photon or electron discussed in the following sections of the
        present paper) and {\it all} path amplitudes consistent with the experimental configuration. 
        \par The fundamental formula (2.4) generalises, by iteration, to give~\cite{JHFAP}:
         \begin{equation}
          P_{FI} =\sum_m \sum_l \left|\sum_{k_n}...\sum_{k_2}\sum_{k_1}\langle f_m|k_n,..k_2,k_1|i_l\rangle\right|^2.
          \end{equation}
         The quantity $P_{FI}$ is the probability to observe any one of the set of final states $F$: $|f_m\rangle$, 
          $m =1,2,3,...$ when any one of the set of initial states $I$: $|i_l\rangle$, $l =1,2,3,...$ is prepared and where
           $|k_1\rangle$,$|k_2\rangle$,...$|k_n\rangle$ are intermediate states that are unobserved, but which
          must be specified in order to calculate the path amplitude:
             \begin{equation}
         \langle f|k_n,...k_2,k_1|i \rangle  \equiv    A_{fi}^{k_n,...k_2,k_1} 
         =  \langle f|k_n\rangle \langle k_n| k_{n-1}\rangle...\langle k_2|k_1\rangle\langle k_1|i\rangle. 
             \end{equation}  
             Three fundamental quantum mechanical laws~\cite{FeynRMP,FH,LLB,JHFAP} are incorporated
             in (2.6):
      \begin{itemize}
            \item[(I)] The Born probability interpretation of the amplitudes: 
               \begin{equation}
                  P_{fi} = |A_{fi}|^2.
               \end{equation}
              \item[(II)] Sequential factorisation of temporally-ordered amplitudes:
             \begin{equation}
                  A_{fi}^k = A_{fk}A_{ki}.
               \end{equation}   
              \item[(III)] Superposition of the path amplitudes $A_{fi}^k$:
           \begin{equation}
          A_{fi} = \sum_k A_{fi}^k = \sum_k  A_{fk}A_{ki}.
         \end{equation}
             \end{itemize} 
         Notice that matrix mutiplication, introduced into quantum mechanics by
         Heisenberg~\cite{HeisMM} and refined by the work of Born and Jordan~\cite{BJ} and by Dirac in his
        `transformation theory'~\cite{DiracTT}, is axiomatically embodied in the laws II and III. 
         \par A fourth important quantum mechanical law is applicable when the initial and final states
            are tensor products such as $|i^{(1)}\rangle\otimes|i^{(2)}\rangle$, $|f^{(1)}\rangle\otimes|f^{(2)}\rangle$.
           Such states occur in experiments where two or more particles are detected in coincidence
            in the final state. Such experiments are conventionally described as having an
           `entangled wavefunction'~\cite{Schrodcat}. In the case that there are no intermediate states
    in common in the
              path amplitudes $\langle f^{(1)}|k_n^{(1)},...k_2^{(1)},k_1^{(1)}|i^{(1)} \rangle$,
            $\langle f^{(2)}|k_n^{(2)},...k_2^{(2)},k_1^{(2)}|i^{(2)} \rangle$, the probability amplitude
            is given by the law:
           \begin{itemize}
            \item[(IV)] Composite factorisation:
             \begin{equation}
             A_{f^{(1)}f^{(2)}i^{(1)}i^{(2)}} =  A_{f^{(1)}i^{(1)}}^{k^{(1)}_n,...k^{(1)}_2,k^{(1)}_1}  
         A_{f^{(2)}i^{(2)}}^{k^{(2)}_n,...k^{(2)}_2,k^{(2)}_1}.
             \end{equation}
   \end{itemize} 
            Since there is no discussion, in the present paper, of experiments with
            entangled wavefunctions, no applications of (2.11) are considered. A brief discussion
            of composite factorisation in the annihilation of para-positronium: $ e^+e^-(^1\rm{S}_0) \rightarrow
             \gamma \gamma$ may be found in Ref.~\cite{JHFAP}.
              \par The laws I-IV above describe simply how probability amplitudes are to be constructed and
               interpreted. They have no dynamical content. Feynman~\cite{FeynRMP}, 
              restricting his discussion to the trajectories
              of particles in space-time, introduced dynamics into the problem, following Dirac~\cite{DiracLQM},
              in the second of the two postulates:
               \begin{itemize}
                \item[A] 
            If an ideal measurement is performed to determine whether a particle has a path lying
            in a region of spacetime, the probability that the result will be affirmative is the absolute
            square of a sum of complex contributions, one from each path in the region.
                \item[B]
            The paths contribute equally in magnitude but the phase of their contribution is the
            classical action (in units of $\hbar$) i.e. the time integral of the Lagrangian taken along the
            path.
                \end{itemize}   
             These postulates give the following law\footnote{Note that this law actually incorporates
            those of sequential factorisation, II, and superposition, III.}
               specifiying the probability amplitude, $A_{BA}$, that 
              a particle, initially, at time $t_A$, at position $\vec{x}_A$,  will be found at later time $t_B$,
              at position $\vec{x}_B$, having followed the spacetime paths: $[\vec{x}^{(j)}(t)]$, $j = 1,2,3,...$:
             \begin{itemize}
            \item[(V)] The Feynman path integral:
              \begin{equation}
               A_{BA} = \sum_j A_{BA}^{(j)} =\sum_j N^{(j)}\exp \left\{ i\frac{S_{BA}([\vec{x}^{(j)}(t)])}
                 {\hbar} \right\} 
                 \end{equation}
                    where
              \begin{equation}
               S_{BA}([\vec{x}^{(j)}(t)]) \equiv \int_{t_A}^{t_B}L([\vec{x}^{(j)}(t)], t)dt.
          \end{equation}
            \end{itemize}
             In (2.12) $ N^{(j)}$ is a (possibly space-time dependent) normalisation factor.
               A particular path  $[\vec{x}(t)^{(j)}]$ is specified by an array
         of space-time coordinates. Considering one spatial dimension:
        \begin{eqnarray}
         &[x(t)^{(j)}] :&~ x_A^{(j)}, t_A^{(j)};~x_1^{(j)}, t_1^{(j)};~x_2^{(j)}, t_2^{(j)};.~.~.
          ~x_n^{(j)}, t_n^{(j)};.~.~.~x_B^{(j)}, t_B^{(j)} \nonumber \\
          &   &  t_A^{(j)}<t_1^{(j)}<~t_2^{(j)}.~.~.<~t_n^{(j)}.~.~.~<~t_B^{(j)}. \nonumber
         \end{eqnarray}
        Notice that the velocity argument, $\dot{x}(t) \equiv d x(t)/dt$, of the Lagrangian is implicit
        in the specification of the path  $[x(t)^{(j)}]$:
         \begin{equation} 
         \dot{x}^{(j)}(t^{(j)}_n) = {\rm Lim}(t^{(j)}_{n+1} \rightarrow t^{(j)}_{n-1})~\frac{{x}^{(j)}_{n+1}-{x}^{(j)}_{n-1}}
               {{t}^{(j)}_{n+1}-{t}^{(j)}_{n-1}}. 
      \end{equation}  
         \par This completes the presentation of the laws of Feynman's formulation of quantum mechanics.
              All the laws are equally valid in both the non-relativistic and relativistic limits. Indeed a prime
             motivation for Dirac's introduction, in (2.12), of the Lagrangian was to enable the construction of
               a relativistic theory, which is not possible in a Hamiltonian-based formulation due to the lack of
              symmetry between temporal and spatial coordinates. 
\SECTION{\bf{The relativistic space-time propagator of a free particle}}
      Writing the exponential factor in Eq.~(2.12) as $\exp(i\phi)$ the phase, $\phi$, is given for a particle of
        Newtonian mass $m$, moving with velocity $\vec{v}$ in free space, by the relations:
         \begin{equation}
      \phi =  \int_{t_A}^{t_B}\frac{L}{\hbar}dt = -\frac{mc^2}{\hbar}\left(\sqrt{1-\beta^2}\right)t \equiv 
          -\frac{mc^2}{\hbar \gamma}t = -\frac{mc^2}{\hbar} \tau = -\frac{P \cdot R}{\hbar} =
            \frac{-Et+\vec{p} \cdot \vec{r}}{\hbar}.
         \end{equation}
       where $\beta \equiv v/c$,  $t \equiv t_B-t_A$, $\tau$ is a proper time interval for the
       particle, $\vec{r} \equiv \vec{x}_B- \vec{x}_A$
         and $R$ and $P$ are space-time and energy-momentum four-vectors of the particle:
         \begin{eqnarray}
          R & = & (R^0,\vec{R}) \equiv (ct,\vec{r}), \\
          P & = & (P^0,\vec{P}) \equiv (\gamma mc,\gamma m \vec{v}) \equiv (\frac{E}{c},\vec{p}), \\
          P \cdot R &\equiv & P^0 R^0-\vec{R} \cdot \vec{P}
           \end{eqnarray}
            where $E$ and $\vec{p}$ are the relativistic energy and momentum of the particle. The expression (3.1)
            incorporates the well-known ~\cite{PlanckRL,GoldRL} non-covariant relativistic Lagrangian:
             $L = -mc^2\sqrt{1-\beta^2}$ for a free particle as well as the relativistic time-dilation relation
             $t = \gamma \tau$. Denoting, following Feynman, the amplitude to find a particle, originally at 
                $\vec{x}_A$ at time $t_A$, at position $\vec{x}_B$ at time $t_B$, the `kernel', or
                `Green's function' or `space-time propagator', by $K(\vec{x}_B,t_B;\vec{x}_A,t_A)$, it follows
                 from (2.12) and (3.1) that:
  \begin{equation}
     K(\vec{x}_B,t_B;\vec{x}_A,t_A) = K(\vec{r},t) = N \exp\left\{-i \frac{(Et-\vec{p} \cdot \vec{r})}{\hbar} \right\}.
       \end{equation}
    The functional dependence of the kernel on $\vec{r}$ is a consequence of translational invariance.
     Because the vectors $\vec{r}$ and $\vec{p}$ are parallel, the phase in (3.1) can be written as
     \begin{equation}
     \hbar \phi = \vec{p} \cdot \vec{r}- Et = pr-Et = p(r-v_{\phi}t)
        \end{equation}
     where a hypothetical, superluminal, `phase velocity':
       \begin{equation}
     v_{\phi} \equiv \frac{E}{p} = \frac{c^2}{v} \ge c
         \end{equation} 
      has been defined by mathematical substitution. The kernel can then be written in a `wave-like'
      manner as:
  \begin{equation}
             K = N\exp i\phi = N f(r-v_{\phi}t)
  \end{equation}                 
    where the last member of (3.8) shows the functional dependence of $K$ on $r$ and $t$. Because of this
      dependence, the kernel satisfies a classical wave equation with phase velocity $v_{\phi}$. 
    The kernel will therefore respect the spherical symmetry of free space  if it satisfies the 
     differential equation:
     \begin{equation}
      \nabla^2 K = \frac{\partial^2 K}{\partial r^2}+\frac{2}{r}\frac{\partial K}{\partial r}
          = \frac{1}{v_{\phi}^2}\frac{\partial^2 K}{\partial t^2}      
 \end{equation}  
       or
    \begin{equation}
       \frac{\partial^2(rK)}{\partial r^2}= \frac{1}{v_{\phi}^2}\frac{\partial^2(rK)}{\partial t^2}
   \end{equation}  
      which has the general solution~\cite{CoulsonWE}:
    \begin{equation}
       K = \frac{1}{r}\left[f(r-v_{\phi}t)+g(r+v_{\phi}t)\right].
      \end{equation}  
     Comparing (3.8) and (3.11) it is seen that
     \begin{equation}
         N =  \frac{1}{r}
      \end{equation}
      so that $K$ has the space-time dependence of a harmonic `spherical wave'.
      \par An alternative derivation of (3.12) is provided by adapting, to the relativitistic
      case, Feynman's  original calculation of normalisation constants in the non-relativistic
      path integral. The kernel $K$ determines the space-time evolution of
     a wavefunction $\psi(\vec{x},t)$ according to the integral equation\footnote{See Ref.~\cite{FH} Eq.~(3.42) p. 57.}:
     \begin{equation}
    \psi(\vec{x}_B,t_B) = \int\int\int K(\vec{x}_B,t_B;\vec{x}_A,t_A)\psi(\vec{x}_A,t_A)d^3x_A.
       \end{equation}
 Setting $t_B = t$, $t_A = t -\epsilon$, where $\epsilon$ is a small fixed increment of time,
      and $\vec{x}_B-\vec{x}_A =\vec{\eta}$, as well as choosing the origin of spatial coordinates
      so that $\vec{x}_B = 0$, gives::
     \begin{equation}
    \psi(0,0,0,t) = \int\int\int K(0,t;-\vec{\eta},t-\epsilon)\psi(-\vec{\eta},t-\epsilon)d^3\eta.
    \end{equation}
    Substituting for the kernel from Eqs.~(3.5) and (3.1) and choosing the $\eta_1$ axis parallel
     to the momentum of the particle gives:
     \begin{equation}
    \psi(0,0,0,t) = \int\int\int N\exp\left\{\frac{i \epsilon}{\hbar}L\left(\frac{|\vec{\eta}|}{\epsilon}\right)
         \right\}\psi(-\vec{\eta},t-\epsilon)
          \delta(\eta_2)\delta(\eta_3)d^3\eta.  
    \end{equation}
 Integrating over $\eta_2$ and $\eta_3$, noting that $\eta_1 = \eta = |\vec{\eta}|$ and substituting $L$ as given
    in (3.1):
     \begin{equation}
    \psi(0,0,0,t) = \int N\exp\left\{-\frac{i \epsilon}{\hbar}m c^2 
     \sqrt{1-\left(\frac{\eta}{c \epsilon}\right)^2}\right\}\psi(-\eta,0,0,t-\epsilon)d\eta.  
    \end{equation}
    Making  Taylor expansions of $\psi(-\eta,0,0,t-\epsilon)$ in the small quantities $\eta$ and $\epsilon$
   and retaining only the zeroth order term gives
     \begin{equation}
    \psi(0,0,0,t) = \int N\exp\left\{-\frac{i \epsilon}{\hbar}m c^2 
     \sqrt{1-\left(\frac{\eta}{c \epsilon}\right)^2}\right\}\psi(0,0,0,t)d\eta.  
    \end{equation}
    Consistency of the zeroth order terms in $\epsilon$ on both sides of this equation requires
  \begin{equation}
       1 = N  \int d\eta = N \eta
    \end{equation}
      or 
   \begin{equation}
         N = \frac{1}{\eta}
       \end{equation}
      in agreement with (3.12) above. 
    \par Combining (3.5) with (3.12) or (3.19) then gives, for the relativistic space-time propagator of a free
       particle of
         mass $m$ and relativistic momentum and energy $\vec{p}$ and $E$, the expression:
       \begin{eqnarray}       
K(\vec{x}_B,t_B;\vec{x}_A,t_A) & = & \frac{1}{|\vec{x}_B-\vec{x}_A|}\exp\left\{-\frac{imc^2}{\hbar}(\tau_B-\tau_A)
            \right\} \nonumber \\
  & = & \frac{1}{|\vec{x}_B-\vec{x}_A|}\exp\left\{-\frac{i}{\hbar}\left[E(t_B-t_A)- \vec{p}\cdot (\vec{x}_B-\vec{x}_A) 
 \right]\right\}.
        \end{eqnarray}
    The phase of the complex exponential in this equation is in agreement with that of the Fourier transform
     of the Lorentz invariant momentum-space propagator, in the limit of large time-like space-time intervals,
     as given in Feynman's first QED paper~\cite{RPFPR}. The $1/r$ factor in the propagator has been given by
     Feynman in an introductory discussion of probability amplitudes\footnote{See
Ref.~\cite{FeynYDS}, Vol III `Quantum Mechanics', Ch. 3 Eq.~(3.7).} as well as the popular book
      `QED'~\cite{FeynQED1}~\footnote{`The {\it length} of the arrow' (i.e. the modulus of the complex probability
       amplitude) `is inversely proportional to the distance the light goes.'\footnote{ See Ref.~\cite{FeynQED1},
    Ch 2 p. 73.} (Feynman's italics)}.
              
\SECTION{\bf{The path amplitude for a photon produced in the decay of an excited atom}}
     As a simple illustration of the laws I and II of Eqs.~(2.8) and (2.9), and the application of the formula
       (3.20) for the space-time propagator of a free particle, consider the probability amplitude to detect a
     photon that orginates in the decay of an excited atom at rest. If the excited state, $i$ \footnote{
       Whether the symbol $i$ denotes a state label or $\sqrt{-1}$ in an equation is clear from
       its context.}, with relativistic
      energy $E_i$, is produced at time $t_0$ and decays at the later time $t_{\gamma}$, into the stable ground
     state of energy $E_f$, the probability amplitude to create the photon may be written as
     $\langle f,t_{\gamma}|i,t_{\gamma}\rangle$. Neglecting the recoil of the
      daughter atom, the time evolution of the initial and final states is given by (3.20) as:
       \begin{eqnarray}
       |i,t_{\gamma}\rangle & = & \exp\left\{-i\frac{(E_i-i\Gamma_i/2)}{\hbar}(t_{\gamma}-t_0)\right\}|i,t_0\rangle \\
        |f,t_{\gamma}\rangle & = & \exp\left\{-i\frac{E_f}{\hbar}(t_{\gamma}-t_0)\right\}|f,t_0\rangle
        \end{eqnarray}
        so that\footnote{See Ref.~\cite{DiracBK}, Section 28 Eq.~(20).}:  
        \begin{equation}
 \langle f,t_{\gamma}|i,t_{\gamma}\rangle = \langle f,t_0|i,t_0\rangle
        \exp\left\{-\frac{i}{\hbar}(E_i-E_f-i\frac{\Gamma_i}{2})(t_{\gamma}-t_0)\right\}
         \end{equation}
          where $\Gamma_i$ is the natural width of the state $i$, related to the mean lifetime $\tau_i$ of the state
       by the energy-time uncertainty relation:
     \begin{equation}
        \Gamma_i = \frac{\hbar}{\tau_i}.
    \end{equation} 
       The factor $\exp[-\Gamma_i/2](t_{\gamma}-t_0)$ in (4.3) takes
       into account the exponential decay law of the excited atom. If the photon is detected at time
        $t_D$ at distance $r$ from the source atom, the photon path amplitude is, from (3.20):
      \begin{equation}
       \langle \gamma, t_D |\gamma, t_{\gamma} \rangle = \frac{1}{r}
        \exp \left\{-\frac{i}{\hbar}[E_{\gamma}(t_D-t_{\gamma})-p_{\gamma}r]\right\}
       \end{equation}
        where
       \begin{equation}
         E_{\gamma} = E_i-E_f.
        \end{equation}
       Since 
   \begin{equation}
  \frac{E_{\gamma}}{p_{\gamma}} = c = \frac{r}{t_D-t_{\gamma}}  
  \end{equation}
     the phase in (4.5) vanishes and
      \begin{equation}
       \langle \gamma, t_D |\gamma, t_{\gamma} \rangle = \frac{1}{r}.
       \end{equation}
     As remarked by Feynman~\cite{FeynQED1}: `Once a photon has been emitted there is no further
     turning of the arrow as the photon goes from one point to another in space-time.' The direction of the
    `arrow' is the phase of the path amplitude. The full probability amplitude, incorporating photon production,
         and propagation, analogous to the `wavefunction' of conventional quantum mechanics, is: 
        \begin{equation}
   \psi_{\gamma}(\vec{r},t_D,t_{\gamma},t_0) = \frac{A_0}{r}\exp \left\{-\frac{i}{\hbar}\left[E_{\gamma}
           -i\frac{\Gamma_i}{2}\right](t_{\gamma}-t_0)\right\}  \langle f, t_0 |i, t_0 \rangle
               \delta[r-c(t_D-t_{\gamma})] 
     \end{equation}
       where $A_0$ is a normalisation constant and the $\delta$-function imposes the space-time
       geometrical constraint relating $r$, $t_D$ and $t_{\gamma}$. The constant  $A_0$ is chosen in 
        such a way that $\psi_{\gamma}$ has the usual Born probabilistic interpretation (Law I):   
              \begin{equation}  
         P_{\gamma}  = \int |\psi_{\gamma}|^2dVdt_D = \int |\psi_{\gamma}|^2d\Omega r^2 dr dt_D  = 1 
      \end{equation}
       where $dV$ is a spatial volume element and $d\Omega$ is an element of solid angle containing the photon path.
       That is, the probability is unity that the photon, once created, can be detected at some position
       at some later time. Notice that the probability amplitude $ \psi_{\gamma}$ is a function,
        not only of the spatial position $\vec{r}$, but also of the times  $t_D$, $t_{\gamma}$
             and  $t_0$ as well as $E_{\gamma}$ and $\Gamma_i$. However, $P_{\gamma}$ can only be normalised,
         as in Eq.~(4.10), by a suitable choice of $A_0$ providing that the the modulus of $\psi_{\gamma}$
         has the spatial dependence $1/r$.
        \par The above example also demonstrates the importance of taking into account all
         relevant physical parameters when analysing a realistic space-time experiment.
         In the example just analysed, the photon travels in a classical manner. The only 
        place where `Heisenberg uncertainty' plays a role is in the energy/time uncertainty relation (4.4). 
         With a typical value of the lifetime of the initial state atom of $\tau_i = 10^{-8}$s the
        momentum/space uncertainty relation: $\Delta x_{\gamma} = \hbar/\Delta p_{\gamma}$ where
         $\Delta p_{\gamma}= \Gamma_i/c = \hbar/(c\tau_i)$ gives  $\Delta x_{\gamma} = 3$ meters.
         A `wavepacket' of this width clearly has no relevance in the simple experiment just described.
         Using a photon detector with a spatial resolution of 10$\mu m$, easily obtained using
         modern solid-state technology, the position of the photon, at the instant of detection, can
         be determined, given the prior knowledge
        of the momentum of the photon, with a precision 33000 times better than `allowed' by the momentum/space
         uncertainty relation! A correct application of this relation would be to the momentum and
        spatial position of an electron in the bound state of an atom. Their distributions are related
        by a Fourier transform which shows that the widths of the distributions are
        related as in the momentum/space uncertainty relation\footnote{This connection between
         Fourier transforms and uncertainty relations was noted at an early date by Bohr~\cite{BohrCI}}.
        In general, no physical significance can be
        attached to the `plane wave' represented by the phase factor
        in Eq.~(3.1) or (3.6), in the absence of the knowledge of other physical parameters necessary to specify
        any realistic space-time experiment where particles travel in free space. In particular, a free
        particle with a known small momentum uncertainty, due to its production process,
        does not have to have a very large uncertainty in position,  as prescribed by the corresponding uncertainty 
       relation.  
    \SECTION{\bf{Reflection diffraction grating for photons}}
\begin{figure}[htbp]
\begin{center}\hspace*{-0.5cm}\mbox{
\epsfysize12.0cm\epsffile{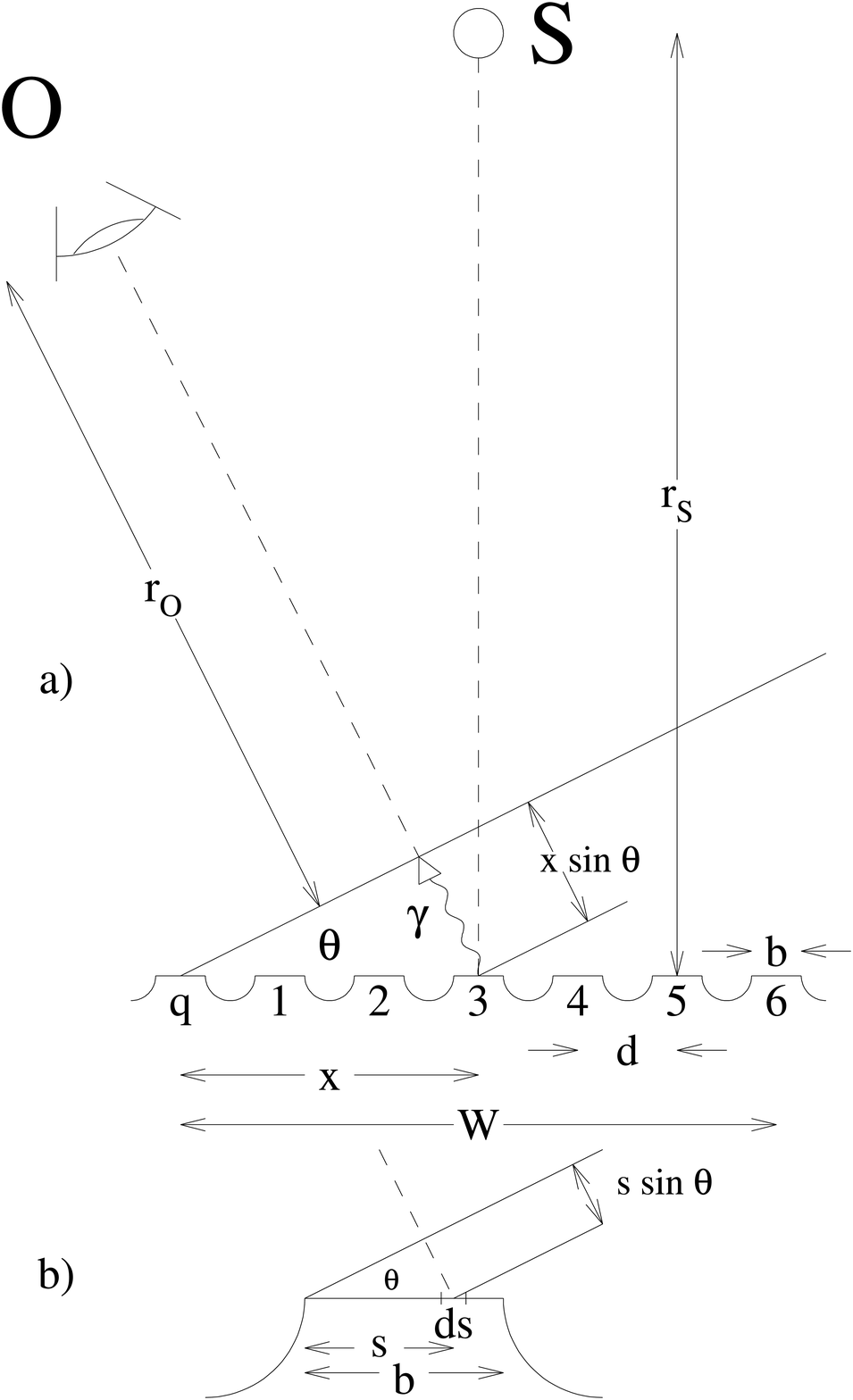}}
\caption{{\em  Geometry of a reflection diffraction grating. A single photon, $\gamma$, is produced in the decay 
            of an excited atom at S, reflected from the grating and observed at O. See text for discussion.}}
\label{fig1} 
\end{center}
\end{figure}
          The experiment discussed in the previous section required only the application of the
          laws I (the the Born probability interpretation) and II (sequential factorisation) as well as the formula
          (3.20) for the space-time propagator of a free particle. The experiments to be discussed in the present
          and following section also bring into play the superposition law III. The geometry of the experiment, 
           which is chosen to be the same as that, using electrons, of Davisson and Germer~\cite{DG}, to be
           discussed in the following section, is shown in Fig.~1\footnote{The experiment can be readily performed,
         in a qualitative manner, by the reader by holding up a compact disc so that light from a localised
         source falls on it. The bright bands, changing rapidly in colour, which are seen as the plane of the
         disc is rotated about a diameter, is the diffraction pattern; the observer's eye serves as the photon
         detector.}. As in the previous section, the probability 
        amplitude for an experiment where a single photon,
         produced by spontaneous decay of a single excited atom, is considered. This atom, constituting the source, S,
         is a distance $r_{{\rm S}}$ from a ruled reflection grating of total width $W$ ($W \ll r_{{\rm S}}$)
         and strip separation $d$. The width of the individual strips is $b$. The $x$ and $y$ axes of a rectangular
         Cartesian coordinate system lie in the
         plane of the grating with the $y$-axis parallel to the strips. The source is centered on the grating
        and lies on the $z$-axis. Since $r_{{\rm S}} \gg W$ all initial photon paths can be considered to be
         parallel
        to the $z$-axis. The plane of the observer O and the source S includes the $z$-axis and is perpendicular
        to the $y$-axis, so that consideration is limited to paths lying in the $x$-$z$ plane. The $N$
        reflecting strips are labelled $0,1,2,...,n$ so that $N = n+1$, the strip $0$ being distant $r_{{\rm O}}$
         $(r_{{\rm O}} \gg W)$ from the observer. The $x$ coordinate of the centre of the $k$th strip is: $x_k =  kd$. 
         \par The excited atom is created at time $t_0$ and subsequently decays at $t_{\gamma}$ ---the creation
          time of the photon. In Fig.~1a is shown the path amplitude of a photon reflected from strip 3
          ---it is shown at the instant after reflection from the strip. The photon is later observed at time
          $t_D$. The initial state of the experiment is that of the excited atom at its instant of creation, $t_0$.
          The final state corresponds to observation (and destruction) of the photon at time  $t_D$.
          The times $t_0$ and  $t_D$ thus label the initial and final states respectively, which are the same
          for all paths considered, like the times $t_A$ and $t_B$ at the limits of the path
         amplitudes discussed in Sections 2 and 3, or the initial and final state labels $i$ and $f$ in
         the superposition law (2.10).
         \par Suitably adapting the formula (4.9), noting the geometry of Fig.~1a, the probability amplitude for a
          photon diffracted from the $k$th 
              strip is:
          \begin{equation}
               \langle t_D|t_{\gamma}|t_0 \rangle = A_0 \AD \frac{1}{r_{{\rm O }}+x_k\sin\theta}  A_{{\rm Diff}}(\theta)
               \frac{1}{r_{{\rm S }}}e^{-\frac{(t_{\gamma}-t_0)}{2 \tau_i}}
              e^{-\frac{i}{\hbar}E_{\gamma}(t_{\gamma}-t_0)}  \langle f, t_0 |i, t_0 \rangle
           \end{equation} 
         where $\AD$ is the amplitude of the photon detection process which includes acceptance factors for solid angles
         subtended at the grating by the source and by the observer at the grating. The energy-time uncertainty
      relation (4.4)
      is used to replace $\Gamma_i$ in (4.9) by $\hbar/\tau_i$. $A_{{\rm Diff}}(\theta)$ is the amplitude
         for diffraction through the angle $\theta$ from any single strip. Assuming
          $x_k\sin\theta \ll r_{{\rm O }}$ and using the space-time geometrical constraint (see Fig.~1a):
           \begin{equation}
           t_{\gamma} = t_D -\frac{r_{{\rm S }}+r_{{\rm O }}+ k d \sin\theta}{c},
              \end{equation} 
           the decay time $t_{\gamma}$ can be eliminated from (5.1) in favour of the strip label $k$ to give:
         \begin{eqnarray}
            A_{FI}^k & \equiv & \langle t_D|t_{\gamma}|t_0 \rangle = A_0 \AD \frac{1}{r_{{\rm O }}}  A_{{\rm Diff}}(\theta) 
                   \frac{1}{r_{{\rm S }}}\exp[i\phi(t_D,t_0)]\exp(i k\alpha) \langle f, t_0 |i, t_0 \rangle
            \nonumber \\ 
             & = & \tilde{A}_0 A_{{\rm Diff}}(\theta) e^{i k\alpha} \equiv \tilde{A}_{0 {\rm D }} e^{i k\alpha}
             \end{eqnarray}
                where
            \begin{equation}
            \phi(t_D,t_0) \equiv -\left(\frac{E_{\gamma}}{\hbar}+\frac{1}{2i \tau_i}\right)
         \left[t_D-\frac{(r_{{\rm S }}+r_{{\rm O }})}{c}-t_0\right]
          \end{equation}
        and  
       \begin{equation}
           \tilde{A}_0 \equiv \frac{A_0 \AD \langle f, t_0 |i, t_0 \rangle}{r_{{\rm O}}r_{{\rm S}}}\exp[i\phi(t_D,t_0)],
        \end{equation}
       \begin{equation}
     \alpha \equiv \left(\frac{E_{\gamma}}{\hbar}+\frac{1}{2i \tau_i}\right)\frac{d \sin \theta}{c} \equiv (r-iq)d
     \equiv \beta d
   \end{equation}
       The superposition law III (Eq.~(2.10)) gives for the probability amplitude of the experiment:
   \begin{equation}
         A_{FI} = \sum_{k=0}^n A_{FI}^k =  \tilde{A}_{0 {\rm D }} \sum_{k=0}^n e^{ik\alpha}
       =   \tilde{A}_{0 {\rm D }}\frac{e^{iN\alpha}-1}{e^{i\alpha}-1}.
   \end{equation} 

     \par Since the geometry of the paths describing reflection of photons from different positions on
       a single strip (Fig.~1b) is similar to that of the paths from different strips of the grating (Fig.~1a)
       the angular dependence of the amplitude $A_{{\rm Diff}}(\theta)$ is given by integrating the paths
       reflected from different positions of the strip. It then follows from the geometry of Fig.~1b that:
       \begin{equation}
    A_{{\rm Diff}}(\theta) = a\int_0^b[\exp(i\beta s)]ds = a \frac{(e^{ib\beta}-1)}{i\beta}
  \end{equation}
    where $a$ is the reflection amplitude per unit width of the strip.
    Combining (5.3), (5.6) and (5.8) the complete probability amplitude for the experiment is:
    \begin{equation}
  A_{FI} = A_0 \AD\frac{1}{r_{{\rm O }}r_{{\rm S }}}    
  \exp[i\phi(t_D,t_0)] a \frac{(e^{ib\beta}-1)}{i\beta}
   \frac{(1-e^{iN\alpha})}{1- e^{i\alpha}} \langle f, t_0 |i, t_0 \rangle.   
   \end{equation}
   Applying the Born interpretation, law I, of Eq.~(2.8) the final probability, $d P_{FI}(t_D)$,
      that the photon will be reflected from
   the grating at angle $\theta$ during the time interval $dt_D$ is then:
     \begin{eqnarray} 
    d P_{FI}(t_D) & = & |A_{FI}|^2 d t_D \nonumber \\
     & = & |\tilde{A}_0 a|^2 | e^{i(\phi(t_D,t_0)-\phi^\ast(t_D,t_0)}
        \frac{(e^{ib \beta}-1)(e^{-ib \beta}-1)}{\beta \beta^\ast}
        \frac{(e^{iN\alpha}-1)(e^{-iN\alpha}-1)}{(e^{i\alpha}-1)(e^{-i\alpha}-1)}d t_D \nonumber \\
         & = & |\tilde{A}_0 a|^2 e^{-\frac{1}{\tau_i}[t_D-\frac{r_{{\rm S}}+r_{{\rm O}}}{c}-t_0]}
           \frac{(1-2e^{qb}\cos br+e^{2qb})}{r^2+ q^2} \frac{(1-2e^{Nqd}\cos Nrd+e^{2Nqd})}{1-2e^{qd}\cos rd+e^{2qd}}d t_D
       \nonumber \\
    & = & \frac{|\tilde{A}_0 a|^2 e^{-\frac{1}{\tau_i}[t_D-\frac{r_{({\rm S}}+r_{{\rm O}}}{c}-t_0]}}{r^2+ q^2}
   \frac{(1+e^{2qb})(1+e^{2Nqd})}{1+e^{2qd}} \nonumber \\
     &  & \times \frac{\left(1-\frac{\cos rb}{{\rm cosh}qb}\right)\left(1-\frac{\cos Nrd}{{\rm cosh} Nqd}\right)}
        {\left(1-\frac{\cos rd}{{\rm cosh}q d}\right)}d t_D.
    \end{eqnarray}
     Considering a typical visual photon energy, $E_{\gamma}$  of 2 eV (e.g. the yellow light of the sodium D-lines) and a natural
     atomic lifetime $\tau_i$ of 10 ns it is found that
      \[ \frac{q}{r} = \frac{\hbar}{\tau_i E_{\gamma}} = 1.64 \times 10^{-8}. \]
      Note that, in virtue of the energy-time uncertainty relation (4.4), the condition $q/r \ll 1$  implies that
     $\Gamma_i  \ll  E_{\gamma}$. Neglecting the imaginary parts of $\alpha$ and $\beta$, since  $q \ll r$,
      and integrating the detection time $t_D$ from its lower limit of $t_0+ (r_{{\rm S }}+r_{{\rm O }})/c$
       to infinity, (5.10) gives:
      \begin{eqnarray}
     P_{FI} & = & 4\tau_i \frac{|\tilde{A}_0 a|^2}{r^2} \frac{\sin^2 \frac{rb}{2}\sin^2 \frac{Nrd}{2}}{\sin^2 \frac{rd}{2}}
           \nonumber \\    
      & = & \tau_i |\tilde{A}_0 a|^2
     \left[\frac{\sin \left(\frac{E_\gamma b \sin \theta}{2 \hbar c}\right)}{\frac{E_\gamma \sin\theta}{2 \hbar c}}\right]^2
     \left[ \frac{\sin \left(\frac{N E_\gamma d \sin \theta}{2 \hbar c}\right)}
       {\sin \left(\frac{E_\gamma d \sin \theta}{2 \hbar c}\right)}\right]^2.
      \end{eqnarray}
     The $\theta$ dependence of this formula, familiar from the classical wave theory of light~\cite{JWDG},
      has here been derived from 
      the path integral formulation of quantum mechanics, without introducing any `wave' concept whatever. 
      \par  Neglecting the imaginary part of $\alpha$ and setting $\alpha = 2 \pi l$, where $l$ is an integer, (5.7) gives:
    \begin{eqnarray}
         A_{FI} & = &  \tilde{A}_{0 {\rm D }}[1 +e^{i2\pi l}+ e^{i4\pi l}+ ...+ e^{i2n \pi l}] \nonumber \\
                & = & \tilde{A}_{0 {\rm D }}[1+1+1+...+1] = N \tilde{A}_0
     \end{eqnarray}
     so that all path amplitudes are in phase and add constructively. According to (5.6) this occurs 
     for all angles $\theta_{{\rm const}}^l$ such that
       \begin{equation}
       \sin \theta_{{\rm const}}^l = \frac{2 \pi l \hbar c}{E_{\gamma} d}~~~~l = 1,2,3,...~.
   \end{equation}
     For $\alpha =(2l+1)\pi$, $l = 0,1,2,...$ (5.7) gives
   \begin{eqnarray}
         A_{FI} & = &  \tilde{A}_{0 {\rm D }}[1 +e^{i(2l+1)\pi}+ e^{i2(2l+1)\pi}+ ...+ e^{in(2l+1) \pi}] \nonumber \\
                & = & \tilde{A}_{0 {\rm D }}[1-1+1-...]  \nonumber \\
                & = & 0~( n{\rm~odd}) \\
                & = &  \tilde{A}_{0 {\rm D }}~(n{\rm~even})
     \end{eqnarray}
         corresponding to destructive interference which is complete for $n$ odd (i.e. for an even number of
         strips)  and partial for $n$ even. This occurs for all angles  $\theta_{{\rm dest}}^l$ such that
        \begin{equation}
       \sin \theta_{{\rm dest}}^l = \frac{(2l+1)\pi \hbar c}{E_{\gamma} d}~~~~l = 0,1,2,...~.
   \end{equation}
     If the first interference maximum occurs for $ \theta = 5^{\circ}$ for a photon energy of 2 eV then (5.12) gives
         \begin{equation}
         d = \frac{hc}{E_{\gamma}\sin 5^{\circ}} = 7.1~\mu{\rm m}. 
      \end{equation}
       For a grating with $b = d$ and 1000 strips (i.e. with a total width of 14.2 mm)
      \[ 2Nqd  = \frac{Nd \sin \theta}{c\tau_i} <  \frac{Nd}{c\tau_i} = 2.4 \times 10^{-3}. \]
      The approximation: $\exp(2Nqd) \simeq 1$ used to obtain (5.11) from (5.10) is therefore good to within  
      a few parts in a thousand for the above choice of diffraction grating parameters and source atom.
       It also follows in this case that damping of the interference pattern by
       the $1/{\rm cosh}Nqd$ factor multiplying the $\cos Nrd$ term in (5.10) is small.

     \par As required by the fundamental law of superposition (2.10), all paths have the same intial and final
      states. In all paths the source atom is produced at the same time $t_0$ and the photon is detected at the same time
      $t_D$. The phases of the different path amplitudes are different because the decay time $t_{\gamma}$
      of the excited atom is different in the different paths, so that each path corresponds to a different
     classical history. This important point was clearly stated in Feynman's last work of popular physics
     ~\cite{FeynQED1} published a quarter of a century ago now. However, this realisation of the
      crucial importance of {\it time} for a correct understanding of quantum mechanical superposition has still 
       not yet penetrated into the relevant research literature\footnote{One isolated example occured in an
       analysis of neutrino oscillations~\cite{KW} where it was stated that: `Since the mass eigenstates propagate
       with different velocities (for fixed energy) the desired interference is between neutrinos emitted at 
       slightly different times.' Actually~\cite{JHFEPJ1} the neutrinos do not have `fixed energy' but certainly 
        have different velocities, so that the statement is a correct one within the path integral formulation.},
        much less into textbooks or the pedagogical literature.
 
 \SECTION{\bf{Reflection diffraction grating for electrons: the Davisson-Germer experiment}}
     The Davisson-Germer experiment~\cite{DG}, performed in 1927, in which the wave-like aspect of massive particles was
      first demonstrated, had the same geometrical configuration as the light diffraction experiment shown in Fig.~1.
       Electrons produced by thermionic emission from a heated tungsten filament
        were accelerated in an electric field. The electron beam thus
      produced struck at normal incidence the [111] face of a face-centered-cubic crystal of nickel atoms with lattice
      spacing 3.51 \AA, The electrons were scattered through an angle $\pi-\theta$ by rows of nickel atoms, with a spacing
      $d =$ 2.15 \AA, that correspond to the reflecting strips in the photon experiment described above. The current of
       scattered electrons was measured by a Faraday box collector over the adjustable angular range
     $20^{\circ} < \theta < 90^{\circ}$. The paths of the incident and scattered beams were of lengths
        $r_{{\rm S}} = 27$mm and $r_{{\rm O}} = 23$mm, respectively. A clear and simple description of the 
       experiment and its results can be found in Ref.~\cite{RKLDGE}.
       \par In this and similar experiments, the electrons are produced not by a well-defined quantum-mechanical
         process as in the photon diffraction experiment described above but by the stochastic process of thermionic
         emission. There is therefore no known phase relation for the amplitudes of electron emission events at
         different times, and the transit time of an electron is determined not, as for a photon, uniquely by the
        length of its path but also by its velocity. Indeed it will be seen that the spread in electron
        velocities is crucial for the observed quantum interference effects. The electrons are emitted from
        the filament with an absolute temperature, $T$, around 2500$^\circ$ K, with a Maxwell-Boltzmann velocity distribution~\cite{JHJ,HHP}
         corresponding to this temperature. The RMS velocity at emission is then
        \begin{equation}
         \bar{v}_{{\rm emit}} = \sqrt{\frac{3kT}{m_{{\rm e}}}} = 1.12 \times 10^{-3}c
         \end{equation}
          where $k$ is the Boltzmann constant and $m_{{\rm e}}$ the mass of the electron.   
         The corresponding kinetic energy is 0.32 eV. For comparison, the electron, after acceleration to a typical
         kinetic energy of 50 eV, has a velocity of $1.4 \times 10^{-2}c$, so that the spread of electron velocities
        is about 6$\%$. As will be shown, this is many orders of magnitude greater than the spread in electron
        velocities that comes into play in the quantum interference effects. 
        \par Unlike the photon experiment, where the photon propagator gives a vanishing contribution to the 
           phase of the path amplitudes, this phase for the electron experiment originates entirely from
          the electron propagator. Writing the propagator phase, $\phi$, as in the first line of Eq.~(3.20)
          and making use of the time dilation relation $t = \gamma \tau$, the relativistic kinematical
          relations $v = pc^2/E$, $E= \gamma mc^2$, and the worldline equation $s = vt$, allows the phase
          to be written as a function
          of only the path length $s$ and the momentum $p$:
        \begin{equation}
      \phi = -\frac{mc^2 \tau}{\hbar} = -\frac{mc^2 t}{\gamma \hbar} = -\frac{(mc^2)^2 s}{ \hbar E v}
            = -\frac{(mc)^2 s}{\hbar p}.
      \end{equation}
        In a similar manner to (5.3) the path amplitude when the electron scatters from the $k$th row of
        nickel atoms is:

         \begin{eqnarray}
            A_{FI}^k & = & A_0 \frac{ A_0 \ADE A_{{\rm Scat}}(\theta) A_{P}^{{\rm e}}}{r_{{\rm S }}r_{{\rm O }}}
       \int f(p_k)\exp[i\phi_k] dp_k 
            \nonumber \\ 
             & \equiv & \tilde{A}_0^{{\rm e}}(\theta)\int f(p_k)\exp[i\phi_k] dp_k
             \end{eqnarray}
        where $A_{P}^{{\rm e}}$ is the production amplitude of the electron,            
    \begin{equation}
\phi_k \equiv -\frac{(mc)^2}{\hbar p_k}s_k = -\frac{(mc)^2}{\hbar p_k}( r_{{\rm S }}+r_{{\rm O }} +kd \sin \theta)
     \end{equation}
    and $f(p_k)$ is the amplitude for production of an electron that has momentum $p_k$ at the scattering event.
    It is distributed around the average momentum $\langle p \rangle$ according to a Gaussian 
    with a width determined by the Maxwell-Boltzmann distribution corresponding to the thermionic emission process:
     \begin{equation}
      f(p) = \frac{1}{(\pi)^{\frac{1}{4}}} \sqrt{\frac{1}{\sigma_p}}\exp\left[-\frac{(p-\langle p \rangle)^2}{2 \sigma_p^2}\right]
     \end{equation}
    where 
      \begin{equation}
        \sigma_p = \sqrt{2 m_{{ \rm e}} k T},~~~T = 2500^{\circ}{\rm K}.
        \end{equation}
       It will be found convenient to discuss first the probability amplitude given by superposition
        of the path amplitudes from two adjacent rows of nickel atoms labelled $k$ and $k+1$:
       \begin{equation}
   A_{FI}^{(2)} = A_{FI}^k+ A_{FI}^{k+1} = \tilde{A}_0^{{\rm e}}(\theta)
        \left[\int f(p_k)\exp[i\phi_k] dp_k+\int f(p_{k+1})\exp[i\phi_{k+1}] dp_{k+1}\right]. 
       \end{equation}
    The corresponding probability, given by the Born interpretation (2.8) is
  \begin{eqnarray}
    P_{FI}^{(2)} & = & |A_{FI}^k+ A_{FI}^{k+1}|^2  \nonumber \\
  & = &  |A_{FI}^k|^2 +|A_{FI}^{k+1}|^2 +2 Re\left[\left(A_{FI}^k\right)^{\ast}A_{FI}^{k+1}\right]
    \end{eqnarray}
   where 
   \begin{eqnarray}
      |A_{FI}^k|^2 & = & |A_{FI}^{k+1}|^2 = |\tilde{A}_0^e|^2\int \int \exp[i(\phi_k(p)-\phi_k(p'))]f(p)f(p')                 \delta(p -p') dp dp' \nonumber \\
             & = & |\tilde{A}_0^e(\theta)|^2
       \end{eqnarray}
      and 
  \begin{eqnarray}
   2 Re\left[\left(A_{FI}^k\right)^{\ast}A_{FI}^{k+1}\right] & = & 2|\tilde{A}_0^e(\theta)|^2 Re\left[ 
     \int \int \exp[i(\phi_{k+1}(p_{k+1})-\phi_k(p_k))] \right. \nonumber \\  
           &   & \times \left. f(p_{k+1})f(p_k)\delta(p_{k+1}-p_k -\Delta p) dp_{k+1} dp_k\right] \nonumber \\  
                     &\equiv & 2 |\tilde{A}_0^e(\theta)|^2 Re (I^{(2)})
    \end{eqnarray}
     where $\Delta p \equiv p_{k+1}- p_k$.  
    \newline From (6.2),
      \begin{equation}
   \Delta \phi \equiv \phi_{k+1}(p_{k+1})-\phi_k(p_k) = \frac{(m_ec)^2}{\hbar}\left(\frac{s_k}{p_k}-
  \frac{s_{k+1}}{p_{k+1}}\right).
     \end{equation}
      The relativistic formula for $\Delta \phi$, correct to first order in $\Delta s/\bar{s}$ where
      $ \Delta s \equiv s_{k+1}- s_k$ and $\bar{s}\equiv (s_{k+1}+ s_k)/2$ is derived in the Appendix:
     \begin{equation}
    \Delta \phi = \frac{(m_e c)^2}{\hbar}\left[ -\frac{\Delta s}{\bar{p}} +\frac{\bar{s}}{\bar{p}}
       \left(1+\frac{\bar{p}^2}{(mc)^2}\right)\left(\frac{\Rc_{{\rm EV}}}{\Rc}-1\right)\right]
       \end{equation}
      where  $\bar{p}\equiv (p_{k+1}+ p_k)/2$,  $\Rc \equiv t_{k+1}/t_k$ and
        $\Rc_{{\rm EV}} =  s_{k+1}/s_k$. Of particular interest are the cases $\Rc = 1$,
          corresponding to equal production times of the electron in the two paths and 
            $\Rc = \Rc_{{\rm EV}}$ which holds for equal velocities (hence the label `EV') in the two paths.
           For $1 < \Rc < \Rc_{{\rm EV}}$ and $ \Rc > \Rc_{{\rm EV}}$  both production times and velocities are different
           in the two paths. It is shown in the Appendix that (6.12) yields the following predictions
           for $ \Delta \phi$ to first order in $\Delta s/\bar{s}$:
         \begin{equation}
        \Delta \phi  = \frac{\bar{p}\Delta s}{\hbar} =\frac{\bar{p} d \sin \theta}{\hbar}~~~
       ({\rm equal~production~times}),  
          \end{equation}
       \begin{equation}
        \Delta \phi  = -\frac{(m_e c)^2 \Delta s}{\hbar \bar{p}} = -\frac{(m_e c)^2  d \sin \theta}
   {\hbar\bar{p}}~~~({\rm equal~velocities}).  
        \end{equation}             
  Considering first the equal production time case and integrating first over $p_k$ gives, for the
     integral $I^{(2)}$ in (6.10):
   \begin{equation}
    I^{(2)} = \frac{1}{\sqrt{\pi} \sigma_p} \int e^{i\frac{(p_{k+1}-\Delta p/2)}{\hbar}\Delta s}
 e^{-\frac{(p_{k+1}-\langle p \rangle)^2}{2 \sigma_p^2}}
   e^{-\frac{(p_{k+1}- \Delta p -\langle p \rangle)^2}{2 \sigma_p^2}}
    dp_{k+1}. 
  \end{equation}
   Performing the integrals over Gaussians in (6.15) by `completing the square'\footnote{With the change
   of variable $p_{k+1}-\langle p \rangle = \tilde{p}$ the integration limits of $\tilde{p}$ are $-\langle p \rangle$
    to infinity, Since $ \sigma_p \ll \langle p \rangle$ the lower limit is set to $-\infty$ in performing the
    integrals in (6.9) and (6.15).}
   the result for $I^{(2)}$ in (6.10) gives:
     \begin{equation}
     2 Re\left[\left(A_{FI}^k\right)^{\ast}A_{FI}^{k+1}\right] =
   |\tilde{A}_0^e(\theta)|^2 e^{-\left(\frac{\Delta p}{2 \sigma_p}\right)^2}
 e^{-\left(\frac{ \sigma_p \Delta s}{2 \hbar}\right)^2}
   \cos\left[\frac{\langle p \rangle}{\hbar} \Delta s\right].
  \end{equation}
    The integrals over the momentum distributions have the effect of replacing the quantity:
        \begin{equation} 
p_{k+1}-\frac{\Delta p}{2} = \frac{p_{k+1}+p_k}{2} = \bar{p}
    \end{equation}
    which is the mean momentum of the particle in the two interfering paths,
     by the quantity $\langle p \rangle$.
    With equal production times:
   \begin{equation} 
        t_{k+1}= t_k =\frac{s_{k+1}}{v_{k+1}} =  \frac{s_k}{v_k}.
     \end{equation}
     Since $\beta \simeq 10^{-2} \ll 1$ in the Davisson-Germer experiment the non-relativistic
     momentum formula,  $p = mv$, holds so that, from (6.18), to first order in $\Delta s/\bar{s}$,
     \begin{equation} 
     \frac{\Delta s}{\bar{s}}= \frac{\Delta v}{\bar{v}} = \frac{\Delta p}{\bar{p}}.
     \end{equation}
      Also, with $T = 2500^{\circ}$ and $\langle p \rangle = 7.2\times 10^{-3}$ MeV/$c$: 
    \begin{equation} 
      \frac{\sigma_p}{\langle p \rangle} =  \frac{\sqrt{2 m_e k T}}{\langle p \rangle} = 6 \times 10^{-2}.
   \end{equation}
      It follows from (6.19) and (6.20) that
   \begin{equation}        
  \frac{\Delta p}{\sigma_p} = \frac{\bar{p}}{\sigma_p}\frac{\Delta s}{\bar{s}}
   \simeq\frac{\langle p \rangle}{\sigma_p}\frac{\Delta s}{\bar{s}} = 16.7\frac{\Delta s}{\bar{s}}.
  \end{equation}
 Because $\Delta s = d \sin \theta < d = 2.15$\AA, the upper limit on $\Delta p/(2 \sigma_p)$ is
   \begin{equation} 
    \frac{\Delta p}{2 \sigma_p} =  8.35 \frac{\Delta s}{\bar{s}}< \frac {8.35 d}{r_{{\rm S}}+ r_{{\rm O}}} = 4.2 \times 10^{-8}.
 \end{equation}
    Since (6.20) gives $\sigma_p = 4.3\times 10^{-4}$ MeV/$c$ then
   \begin{equation}  
       \frac{\sigma_p \Delta s}{2 \hbar} = \frac{\sigma_p d \sin \theta}{2 \hbar}
    < \frac{\sigma_p d}{2 \hbar} = 8.9 \times 10^{-9}. 
    \end{equation}
     Finally (6.20) and (6.22) give
     \begin{equation}  
       \frac{\Delta p}{\bar{p}} \simeq \frac{\Delta p}{\langle p \rangle} < 5.0 \times 10^{-9}.
   \end{equation}
     The momentum difference $\Delta p$ that is the physical basis of the interference of the path amplitudes
     of electrons scattering from adjacent rows of atoms is therefore seven orders of magnitude smaller than
     the momentum spread of the beam in the experiment. 
    \par The left sides of (6.22) and (6.23) appear squared as the arguments of the negative
     exponential factors in (6.16). The interference damping given by these factors it therefore
     completely negligible. Setting therefore
   the damping factors to unity, (6.8), (6.10) and (6.16) may be combined to give: 
    \begin{equation}
    P_{FI}^{(2)}  = |A_{FI}^k+ A_{FI}^{k+1}|^2 
     =  |\tilde{A}_0^e(\theta)|^2\left[1 + 
  \cos\left( \frac{\langle p \rangle d \sin \theta}  {\hbar} \right) \right]~~
       ({\rm equal~production~times}).  
    \end{equation}
     Notice the important point that, although the damping produced by the factor \newline
     $\exp[-(\Delta p/2 \sigma_p)^2]$ in (6.16) is completely negligible, this factor, and hence
     the interference term, vanishes for vanishing $\sigma_p$ for any finite value of $\Delta p$.
     The interference effect therefore requires different momenta in the two paths for equal
     production times in the paths. The factor  $\exp[-(\sigma_p \Delta s/2 \hbar)^2]$ in (6.16)
     shows that the interference effect is also destroyed for large values of the path difference
     $\Delta s \gg 2 \hbar/\sigma_p$, i.e. for values such that $\Delta s$  is much larger than the
     width of a hypothetical spatial electron `wave packet' given by the Fourier transform of the
     amplitude $f(p)$ in Eq.~(6.5).
      \par For the equal velocity case, the $\delta$-function $\delta(p_{k+1}-p_k - \Delta p)$ 
      in (6.10) is replaced by  $\delta(p_{k+1}-p_k)$ so that that the integrals over electron
      momenta are similar to that in (6.9) and there are no damping factors containing $\sigma_p$.
      Also, since the effect of integration over the electron momentum distributions is to 
      replace $\bar{p}$ in the interference term by $\langle p \rangle$ (see Eqs.~(6.15), (6.16) 
      and (6.17)) the probability distribution for equal velocities is:
     \begin{equation}
    P_{FI}^{(2)} = |\tilde{A}_0^e(\theta)|^2 \left[1 + 
  \cos\left( \frac{(m_{{\rm e}}c)^2 d \sin \theta}  {\hbar\langle p \rangle } \right) \right]~~
       ({\rm equal~velocities}).  
      \end{equation}
       \par With the definition:
 \begin{equation}
             \alpha_{{\rm e}} \equiv \frac{\langle p \rangle d \sin \theta}{\hbar}
     \end{equation}
      then, since
 \begin{equation}
          \frac{\sin^2 2\left(\frac{\alpha_{{\rm e}}}{2}\right)}{\sin^2\frac{\alpha_{{\rm e}}}{2}}
         =  \frac{4 \sin^2\frac{\alpha_{{\rm e}}}{2}\cos^2\frac{\alpha_{{\rm e}}}{2}}
             {\sin^2\frac{\alpha_{{\rm e}}}{2}}
         = 4\cos^2\frac{\alpha_{{\rm e}}}{2} = 2(1+\cos\alpha_{{\rm e}},).
    \end{equation}
        the equal production time formula (6.25) has the same $\theta$ dependence
        as given by the replacements: $N = 2$, $E_{\gamma}/c =  p_{\gamma} \rightarrow \langle p \rangle$
        in the last factor of the equal velocity, different production time, formula (5.15) for
        the photon experiment. The quantity $P_{FI}^{(N)}$ for $N> 2$ is therefore also expected
        to be proportional to $\sin^2(N\alpha_{{\rm e}}/2)/\sin^2(\alpha_{{\rm e}}/2)$. Since the
       $\sigma_p$ dependent damping factors may be set to unity, this result may be derived by
        introducing momentum-averaged phases for the paths according to the
        equations:
        \begin{eqnarray}
        \langle \phi_k^{{\rm EPT}}\rangle & = & \frac{\langle p \rangle s_k}{\hbar}
     = k \alpha_{{\rm e}}^{{\rm EPT}} + \phi_0^{{\rm EPT}}~~({\rm equal~production~times}),  \\ 
   \langle \phi_k^{{\rm EV}}\rangle & = & -\frac{(m_{{\rm e}}c)^2 d \sin \theta}
      {\hbar\langle p \rangle } s_k = k \alpha_{{\rm e}}^{{\rm EV}}+ \phi_0^{{\rm EV}}~~
       ({\rm equal~velocities})  
        \end{eqnarray}
         where
   \begin{eqnarray}
  \alpha_{{\rm e}}^{{\rm EPT}} &\equiv & \frac{\langle p \rangle d \sin \theta}{\hbar},~~~~
 \alpha_{{\rm e}}^{{\rm EV}} \equiv  -\frac{(m_{{\rm e}}c)^2 d \sin \theta} {\hbar\langle p \rangle}, \\
   \phi_0^{{\rm EPT}} &\equiv & \frac{\langle p \rangle}{\hbar}(r_{{\rm S}}+ r_{{\rm O}}),~~~~
    \phi_0^{{\rm EV}} \equiv 
   -\frac{(m_{{\rm e}}c)^2 d \sin \theta} {\hbar\langle p \rangle}(r_{{\rm S}}+ r_{{\rm O}}).
    \end{eqnarray}
    Eq.~(6.7) then generalises to:
   \begin{eqnarray}
      A_{FI}^{(N)} & = &  \tilde{A}_0^e(\theta) \left[e^{i \langle \phi_0 \rangle}+ e^{i \langle \phi_1 \rangle}
      +...+e^{i \langle \phi_{N-1} \rangle} \right] \nonumber \\
         & = &  \tilde{A}_0^e(\theta) \left[1+  e^{i \alpha_{{\rm e}}}+  e^{i 2 \alpha_{{\rm e}}}
      +...+ e^{i(N-1) \alpha_{{\rm e}}}\right]e^{i\phi_0} \nonumber \\
          & = &  \tilde{A}_0^e(\theta) e^{i\phi_0}  \frac{(1- e^{i N \alpha_{{\rm e}}})}{1- e^{i\alpha_{{\rm e}}}}  
  \end{eqnarray}
    so that the formula for the angular distribution of an electron diffracted from $N$ rows of
     nickel atoms in the Davisson-Germer experiment is:
       \begin{equation}
        P_{FI}^{(N)}(\theta) = |A_{FI}^{(N)}|^2 = |\tilde{A}_0^e(\theta)|^2 
       \frac{\sin^2 N\left(\frac{\alpha_{{\rm e}}}{2}\right)}{\sin^2\frac{\alpha_{{\rm e}}}{2}} 
     \end{equation}
      where $\alpha_{{\rm e}} = \alpha_{{\rm e}}^{{\rm EPT}}$ or  $\alpha_{{\rm e}}^{{\rm EV}}$.
      The first diffraction maximum occurs when $|\alpha_{{\rm e}}| = 2 \pi$ giving the predictions:
    \begin{eqnarray}
     \sin \theta_{{\rm max},1}^{{\rm EPT}} & = & \frac{2 \pi \hbar}{\langle p \rangle d}   
     ~~({\rm equal~production~times}),  \\ 
     \sin \theta_{{\rm max},1}^{{\rm EV}} & = & \frac{2 \pi \hbar \langle p \rangle }{(m_{{\rm e}}c)^2 d}~~
       ({\rm equal~velocities}).  
       \end{eqnarray}
    Substituting $\langle p \rangle = 7.43\times 10^{-3}$MeV/$c$ corresponding to $T_e = 54$ eV and
    $d = 2.15$\AA~ in these formulas gives $\theta_{{\rm max},1}^{{\rm EPT}} = 51^{\circ}$ in good agreement
     with the observation of the Davisson-Germer experimemt~\cite{DG,RKLDGE} whereas
    $\theta_{{\rm max},1}^{{\rm EV}} = 0.0094^{\circ}$. Only the equal production time hypothesis is therefore
     consistent with the experiment. A consequence is that the prediction does not depend on any phase, with a possibly
       stochastic time dependence, of the electron production amplitude $A_P^{{\rm e}}$, which is not the case
      for equal velocities and different production times.
     \par The above analysis neglects the angular dependence of the scattering amplitude $ A_{{\rm Scat}}(\theta)$
    of an electron on a single nickel atom, which can only be obtained by performing the appropriate
     quantum-mechanical calculation. For the non-relativistic electrons in the Davisson-Germer experiment
     the corresponding angular distribution is expected to be isotropic, or in any case, much less rapidly-varying
     than the single-strip diffraction factor for photons in (6.34). The approximation of assuming an 
    angle-independent scattering amplitude in calculating the shape of the overall diffraction
     pattern is therefore expected to be a good one. 

    \SECTION{\bf{Classical wave theories of photons and massive particles}}
         Using Eq.~(5.2) the photon path amplitude of Eq.~(5.3) may be split into constant, time-dependent
         and spatially-dependent factors as:
         \begin{eqnarray}
    A_{FI}^k & = & \frac{A_0 \AD  A_{{\rm Diff}}(\theta) \langle f, t_0 |i, t_0 \rangle}{ r_k r_{{\rm S }}}
           \nonumber \\
         &\times & \exp\left[-\frac{i}{\hbar} \left(E_{\gamma}-i\frac{\Gamma_i}{2}\right)(t_D-t_0)\right]
            \nonumber \\
        & \times & \exp\left[\frac{i}{\hbar}\left(E_{\gamma}-i\frac{\Gamma_i}{2}\right)\frac{(r_k+r_{{\rm S }})}
         {c}\right]
   \end{eqnarray}
    where
    \begin{equation}
        r_k \equiv r_{{\rm O }}+kd\sin \theta.
     \end{equation}
     The $1/r_k$ dependence of (7.1), unlike the $1/r_{{\rm O }}$ dependence of (5.3) is exact to first
       order in $\Delta s/\bar{s}$. Consider now observation of the photon at different positions for
       a fixed source position. Then $r_{{\rm S }}$, $t_D$ and $t_0$ are constants for different paths
        labelled by $k$ so that $A_{FI}^k$ is a function only of $r_k$:
    \begin{equation}
      A_{FI}^k \rightarrow U_{\gamma}(r_k) \equiv \frac{(\tilde{A}_0^{{\rm S}})_{\gamma}}{r_k}  
  \exp \left[\frac{i}{\hbar c}\left(E_{\gamma}-i\frac{\Gamma_i}{2}\right)r_k\right]
 \end{equation}
   where the label S in the constant amplitude $(\tilde{A}_0^{{\rm S}})_{\gamma}$ stands for
   `source'. If $\Gamma_i/(2\hbar c)r_k \ll 1$, so that damping effects due to the finite lifetime of the
       source atom are negligible, (7.3) simplifies to:
 \begin{equation}
      U_{\gamma}(r_k) = \frac{(\tilde{A}_0^{{\rm S}})_{\gamma}}{r_k}  
  \exp \left[\frac{i E_{\gamma}r_k}{\hbar c}\right]= \frac{(\tilde{A}_0^{{\rm S}})_{\gamma}}{r_k}  
  \exp \left[\frac{i 2 \pi r_k}{\lambda_{\gamma}}\right]
 \end{equation}
   where
    \begin{equation}
   \frac{E_{\gamma}}{\hbar c} \equiv \frac{2 \pi}{\lambda_{\gamma}} \equiv \kappa_{\gamma}.
   \end{equation}
 The quantity $\lambda_{\gamma} = hc/E_{\gamma} = h/p_{\gamma}$ is the `de Broglie wavelength' of the
 photon. However the discussion in Section 5 above shows that the phase in (7.4) originates entirely 
  from the time dependence of the decay amplitude of the unstable atom (Eq.~(4.3)) and so is in no
  sense an attribute of the photon itself. Eq.~(7.4) shows that the photon path amplitude is equivalent
  to the spatial part of a classical wave with phase velocity $c$ and wavelength  $\lambda_{\gamma}$.
  The relation $c = \lambda_{\gamma} \nu$ follows from (7.5) and the Planck-Einstein relation for photons: $E_{\gamma} = h \nu$.
   All predictions of interference and diffraction
  effects  for light in the case that the lifetime of the excited atom can be considered to be infinitely long
  can therefore be obtained from (7.4) in which only spatially dependent waves are considered. This
  is the 19th Century `classical wave theory of light'. The theory is `classical' because Planck's constant
     does not appear in the equations, its effect being hidden in the phenomenological photon wavelength
     parameter. The temporal physical origin of the phase, evident in
  Eq.~(4.3), is transformed away when $t_{\gamma}$ is eliminated from (5.3) using (5.2) to yield (7.1).
 \par Considering, for electrons, the equal production time case, as experimentally verified in the
   Davisson-Germer experiment, the path amplitude formula analogous to (7.3) is:   
   \begin{eqnarray}
  A_{FI}^k & = & \frac{A_0 \ADE A_{{\rm Scat}}(\theta) A_P^{{\rm e}}}{ r_k r_{{\rm S }}}
              \exp \left[i\frac{\langle p \rangle}{\hbar}(r_k+r_{{\rm S }})\right]~~\rightarrow \nonumber \\
     U_{{\rm e}}(r_k) & \equiv  & \frac{(\tilde{A}_0^{{\rm S}})_{{\rm e}}}{r_k}  
  \exp \left[\frac{i \langle p \rangle r_k}{\hbar}\right] = \frac{(\tilde{A}_0^{{\rm S}})_{{\rm e}}}{r_k}  
  \exp \left[\frac{i 2 \pi r_k}{\lambda_{{\rm e}}}\right]
 \end{eqnarray}
    where the `de Broglie wavelength' of the electron:
   $\lambda_{{\rm e}} \equiv h/\langle p \rangle\equiv 2 \pi/\kappa_{{\rm e}}$,  although defined in a similar way, as
   $h/p$, is, unlike that of the photon, an attribute of the electron originating in its space-time
   propagator. There is no time dependence in the path amplitude (7.6) as a consequence of the equal time
   condition for different paths (i.e. for different values of $k$). The formula (7.6) gives a classical
   theory of `matter waves' strictly analogous to the classical wave theory of light.    
    \par The spherical spatial waves of (7.4) or (7.6) are solutions of the Helmholtz equation:
          \begin{equation}
          \nabla^2U +\kappa^2U = 0
           \end{equation}
          which serves as the basis of classical wave theories of both light and material particles,
     even though the underlying space-time physics is quite different in the two cases. As a consequence
     of (7.7), diffraction is described by  Kirchoff's equation~\cite{BW}, and Huygen's construction
     can be used to perform a purely spatial analysis of wavefronts in conjuction with the phenomenological
     wavelength parameters. Planck's constant is thereby banished from all equations and the purely quantum
     mechanical nature of the fundamental underlying physics is hidden. Indeed the simple classical wave formalism
     of Eq.~(7.4) gives many quantitatively correct results in physical optics and the classical wave
     formula Eq.~(7.6) was used in the 
     original interpretion of the Davisson-Germer experiment, to correctly predict the observed
      diffraction effects. Historically, of course, the discovery
     of the phenomenological wave theory of light, to explain the interference experiments of Fresnel and Young,
     predated by a century that of quantum mechanics and lead to the false
     ontological identification of light with classical waves, i.e. as a disturbance of some material medium.This
     misidentification was reinforced by the advent of Maxwell's electromagnetic wave theory of light and the
     associated models of a material luminiferous aether to support the putative wave motion.
     After a century of experimental particle physics it is now known, beyond reasonable doubt, that
     both photons and electrons 
     are indeed particles in the ontological sense. The temporal aspects
     of their motion in space-time, inexplicable by classical wave theory, may be clearly demonstrated by
     considering suitably chosen 
     experiments. An example of this is the damping effect of the $1/{\rm cosh}$ factors in Eq.~(5.10) reflecting the 
     finite lifetime of the source atom.  Some other examples may be found in Ref.~\cite{JHFAP}.

\SECTION{\bf{Conclusions for the physical interpretation of quantum mechanics}}

       Some remarks are now made concerning the `interpretation of quantum mechanics'
       in the context of Feynman's space path integral formulation and in the light of the space-time
       analyses of the experiments presented in Sections 4, 5 and 6 above. In typical text book presentations
       of quantum mechanics the primary physical concept is not, as in Feynman's formulation, the 
        probability amplitude $A_{fi}$ but the wavefunction
         $\psi(\vec{x}_1,\vec{x}_2,...\vec{x}_N;t_1,t_2,...t_N)$ of a quantum mechanical system composed of $N$ 
      particles. For a single particle, the wavefunction is assumed to be a function of only of the space-time coordinates of
     this particle and so constitutes a `field'. In the non-relativistic theory the properties of the wavefunction
    are assumed to be those of a solution of the Schr\"{o}dinger equation. In contrast, as demonstrated by the
     calculations presented in Sections 4, 5 and 6 above, the probability amplitude for a quantum experiment depends
    also on the space-time coordinates of the source of particles considered as well as those of the particle (or particles)
     which are detected  in the experiment.
     \par In the `Copenhagen Interpretation' of quantum mechanics~\cite{HeisCI,BohrCI}, which is the one adopted
    by essentially all text books on the subject, as well as in the pedagogical literature and many popular accounts,
     there is supposed to be an ontological dualism between `waves' and `particles', termed `wave-particle duality'.
     It is commonly stated that an electron `sometimes behaves like a wave' and `sometimes behaves like a particle'.
     This begs the obvious question: What {\it is} an electron, is it a particle or is it a wave? 
     Every particle physicist who has ever built or worked on an actual experiment knows, beyond any
     doubt, that, operationally speaking, the electron (and the photon) {\it are} particles in the same sense that this word was
     understood by Newton. This is the essential concept required to design and understand what are
     correctly called `particle physics' experiments. Even experiments such as those described in Sections 5 and 6
      above where the `wave like' properties of particles are demonstrated are analysed, in Feynman's
     formulation of quantum mechanics, entirely in terms of `what happens' to entities localised in time and
     space and moving classically ---particles.
     \par The formulas of the path integral formulation contain only space-time coordinates and kinematical
      quantities (energies, momenta and masses) of {\it particles} so the question  of  `wave-particle duality'
      cannot even arise in `interpretations' of this formulation. So where do the `waves' come from in the
       traditional Schr\"{o}dinger formulation? By formal mathematical substitution it is possible
     to eliminate Planck's constant, $h$, completely from all equations, in favour of the de Broglie wave length, $\lambda$,
     by using the relation $h = p\lambda$. This leads to the `wave-like' space-time functional
      dependence of the free particle propagator as in Eq.~(3.8) with a superluminal `phase velocity':
       $v_{\phi} =c^2/v$. By introducing a `refractive index' of free space: $n \equiv v/v_{\phi}$ and using Lord Rayleigh's
       group velocity formula: $1/u = (1/c)d(n\nu)/d\nu$ de Broglie showed~\cite{deBPhD,DBNPS,JHFHJE} that $u = v$ i.e. that the
        particle velocity is equal to the group velocity of a packet of `phase waves' with average velocity $v_{\phi}$.
        The same conclusion may be drawn by defining the group velocity as $u \equiv dE/dP$ and using the relativistic
        formula relating energy, momentum and mass together with the Planck-Einstein relation
        $E = h \nu$~\cite{JHFHJE}. A consequnce of this purely mathematical manipulation is that many text books
       state, in an introductory chapter, that a `particle' in quantum mechanics  {\it is} a packet of phase
        waves with widths in momentum space $\Delta p$ and physical space $\Delta x$ that satisfy the Heisenberg
        Uncertainty relation $\Delta p \Delta x = \hbar$. In the analysis of the photon experiments described in Sections
        4 and 5 above no such `wave packet' occurs. This identification of particles with wave packets
       is part of Bohr's original specification of the `Copenhagen Interpretation'~\cite{BohrCI}:
      \par{\tt The circumstance that the}~(phase velocity)~{\tt is in general greater than the \newline
    velocity of light
    emphasises the symbolic character of these considerations. At the same time the possiblilty of indentifying 
     the velocity of the particle with the group velocity indicates the field of applicability of
       space-time pictures in the quantum theory.}
      \par  The photon propagates in space time, within each path, not as a `wave packet' but as a classical
       particle with fixed momentum and a constant velocity\footnote{This is only the case in `flat space' where
     all gravitational effects are negligible.} that is the same in all paths. In the Davisson-Germer
      experiment, momentum wave packets do occur in the electron path amplitudes in virtue of the thermionic emission
      process that liberates the electrons. The electrons then propagate, with a constant, but path-dependent, velocity
      in each path. The corresponding spatial wave packet, derived from Eq.~(6.5) by Fourier transformation will
      have, as first pointed out by Bohr~\cite{BohrCI}, a width $\sigma_x$ that respects the uncertainty-like
     relation $\sigma_p \sigma_x = \hbar$, but this hypothetical spatial wave packet, unlike the
       Maxwellian momentum distribution, plays no role in the
      space-time analysis of the Davisson-Germer experiment. The only relevant application of a Heisenberg
      Uncertainty relation in the experiments described above is the use of energy-time relation (4.4),
     which connects the decay width of the excited state to its mean lifetime. As explained at the end of Section 5,
    there is no corresponding photon wave packet, and the position and momentum of the photon can both be known with
    a precision much greater than allowed by a putative momentum-space uncertainty relation with
     $\Delta p_{\gamma} = \Gamma_i/c$.
     \par To make clear the different ontological nature of particles and the `waves' of wave mechanics or the probability
      amplitudes of Feynman's formulation, it is instructive to consider first a typical system described by 
     classical mechanics. It consists of a part that exists in the real world (e.g. the Earth-Moon system) and is
      {\it described by}  
     an abstract mathematical entity (e.g. a Lagrangian) from which its space-time behaviour (its motion)
     can be derived. For a quantum mechanical system, say a particle moving in a region of known
     potential energy, there is also a part that exists in the real world 
     (the particle and the source of the potential)
     as well as the probability amplitude, which is a useful mathematical abstraction,
     from which statistical information
     on the space-time evolution of the system can be derived. It turns out that, in Feynman's formulation,
      the essential ingredient of the probability amplitude is, according to the law V, the same as for the
      classical example, the (classical) Lagrangian function of the system. In fact as pointed out
     by Dirac~\cite{DiracLQM},  the classical motion is recovered from the probability amplitude in
     the limit $h \ll A_{{ \rm min}}$ where $A_{{ \rm min}}$ is the minimum size of any quantity 
     with dimensions of action that enters into the physical description of the system under consideration.
      The ontological confusion of statements
     like `an electron sometimes behaves like a particle, and sometimes behaves like a Lagrangian' or `an electron is
    sometimes a particle and sometimes a Lagrangian' is quite evident. In fact an electron {\it is} always a particle
    and its motion in space-time is always {\it described} by an appropriate (classical) Lagrangian in both classical
    and quantum mechanics in the space-time formulation due to Dirac and Feynman.  Quite simply, in this case,
    there is no ontological duality.
    \par The failure of the matrix mechanics and wave mechanics formulations to incorporate space-time concepts
           in a transparent manner, as well as the ontological confusion between `particles' and `waves' which has
     persisted until the time of this writing is primarily due to two circumstances: (1) Because of the quantitative
     success of the 19th century classical wave theory of light as developed by Young, Fresnel and others, as well
     as Maxwell's identification of light with electromagnetic waves, the concept of `light waves' was already firmly
     woven into the fabric of physics at the time of the advent of quantum mechanics.(2) The problem which was
     addressed by the pioneers of quantum mechanics ---the theory of atomic structure and atomic radiative
     transitions--- was one one in which space-time concepts play no essential role. As correctly remarked
     by Bohr~\cite{BohrCI}:
     \par{\tt For example, the experiments regarding the excitation of spectra by \newline electronic impacts
              and by radiation are adequately accounted for on the \newline assumption of discrete stationary
               states and individual transition processes. This is primarily due to the
              circumstance that in these questions  no closer description of the space-time behaviour 
               of the processes is required.}
        \par Indeed, the wavefunction concept and the Schr\"{o}dinger equation are essential for the quantum
             description of atomic states and atomic transitions (which may be characterised as
             `quantum statics') but they are, unlike the probability amplitude
           of Feynman's formulation, ill-adapted to the description of dynamical experiments where space-time ideas 
        are primordial, such as those described in Sections 4, 5 and 6 above or~\cite{JHFAP}.
          \par Another subject to be found in typical quantum mechanical text books is `measurement theory' based on 
        projection operators in an abstract Hilbert space, following the methodology introduced by von Neumann
        in the early 1930's~\cite{VN}. `Measurement' is defined as the projection of an eigenfunction
        out of some general  Hilbert space state vector. For example\footnote{See Ref.~\cite{DiracBK}, Sect. 10, p. 36.}:
     \par{\tt ...we see that a measurement always causes a system to jump into an \newline eigenstate
              of the dynamical variable that that is being measured, the\newline  eigenvalue of this eigenstate 
             being equal to the result of the measurement.}
       \par In the case of the experiments described in Ref.~\cite{JHFAP} and in Sections 4 and 5 above `measurement' (i.e. detection of
         the photon) does not produce a photon in `an eigenstate of position' but simply destroys it.
           The electrons which are detected in the Faraday box of the Davisson-Germer experiment
            are also not left in `position eigenstates' after detection. To state the situation bluntly,
            the basic concepts of text book quantum mechanical `measurement theory', although mathematically
            elegant, have little direct applicability to most real-world experiments. Rigorous conclusions are obtained
            within a limited, well-defined, but abstract model. However this model is, in general, too simplistic
           to address the complexity of actual, real world, experiments like those considered in Sections 5 and 6
           above. 
           \par As a last application of the path integral formalism, consider the well-known `Schr\"{o}dinger's cat'
              thought experiment of wave mechanics~\cite{Schrodcat}:
            \par{\tt A cat is penned up in a steel chamber along with the following diabolical device (which must
                  be secured against direct influence by the cat): in a \newline Geiger counter there is
               a tiny bit of radioactive substance so small that}\newline~{\it perhaps}~{\tt in the course of one hour
               one of the atoms decays, but also, with equal probability, perhaps none; if it happens, the counter
                tube discharges and \newline through a relay releases a small hammer which shatters a small flask of
                 \newline hydrocyanic acid. If one has left this entire system to itself for an hour one would say
                that the cat still lives}~{\it if}~{\tt meanwhile no atom has decayed.\newline The first atomic decay 
                would have poisoned it. The $\psi$  function of the \newline entire system would express this by having
                in it the living and dead cat  \newline (pardon the expression) mixed, or smeared out in equal parts.}
                 (italics in the orginal)
              \par  The conditions of the experiment are met by considering a single radioactive atom of mean
                 lifetime $\tau = 1.44$~hr. What is relevant here is not `The $\psi$  function of the entire system'
                  but the time-dependent transition amplitude: $\langle f,t|i,t \rangle$ of the decay process where
                  $i$ and $f$ are the initial and final states of the unstable radioactive nucleus. Using (4.9) and (2.8) the probability that the cat is 
                    still alive after a time interval $t$ is:
                    \begin{equation}
                    P({\rm cat~alive}) = \frac{|\langle f,t-t_{{\rm Del}}|i,t-t_{{\rm Del}}\rangle|^2}{|\langle f,0|i,0 \rangle|^2}
                          = \exp\left\{-\frac{(t-t_{{\rm Del}})}{\tau}\right\}
                        \end{equation} 
                       where $t_{{\rm Del}}$ is the delay time, determined by a chain of events related by classical causality, after 
                      the decay of the radioactive atom, before
                      the cat actually dies. The cat is dead at, or after, the time interval $t$ if the atom
                      decays within
                      the time interval $t-t_{{\rm Del}}$ so that:
                    \begin{equation}
                    P({\rm cat~dead}) =1- P({\rm cat~alive})= 1- \exp\left\{-\frac{(t-t_{{\rm Del}})}{\tau}\right\}
                      \end{equation} 
                    since the atom either decays, or does not decay, within the time interval $t-t_{{\rm Del}}$.
                     The probabilities in (8.1) and (8.2) are independent of whether the chamber is opened 
                     or remains shut after the experiment, i.e. whether the cat is observable or not.
                     At each instant during the experiment the cat
                     is either alive or dead so its wavefunction is either $\psi({\rm cat~alive})$
                     or $\psi({\rm cat~dead})$. Schr\"{o}dinger's putative wavefunction, after one hour,
                       (neglecting the delay $t_{{\rm Del}}$) of
                     $\psi = [\psi({\rm cat~alive}) +\psi({\rm cat~dead})]/\sqrt{2}$
                     therefore has no relevance to the analysis of the experiment. Note that the quantum mechanical
                     superposition evident in Schr\"{o}dinger's  wavefunction plays no role in the derivation
                     of the prediction (8.2) in Feynman's formulation, much less `entangled' wavefunctions, that
                     are appropriate only to quantum experiments 
                     with dual final states.
 \pagebreak 
\par {\bf Appendix}  
\renewcommand{\theequation}{A.\arabic{equation}}
\setcounter{equation}{0}
 \par Consider paths A or B followed by a free particle of mass $m$ with velocities $v_A$ or $v_B$.
   If the corresponding times of transit are $t_A$, $t_B$ then the lengths of the paths are:
   $s_A = v_A t_A$,  $s_B = v_B t_B$. With the definitions:
     \begin{equation}
  \Rc \equiv \frac{t_A}{t_B},~~~~ \Dc \equiv t_A-t_B
      \end{equation}
        it follows that
       \begin{equation}
      v_A = \frac{(\Rc-1)s_A}{\Rc\Dc},~~~ v_B = \frac{(\Rc-1)s_B}{\Dc}.
        \end{equation}
 If $t_A \ge t_B$ then the physically allowed regions of $\Rc$ and $\Dc$ are:
     \begin{equation} 
          1 \le \Rc < \infty,~~~~~0 \le \Dc < \infty.
       \end{equation}
  The conditions $\Rc = 1$ and  $\Dc = 0$ correspond to equal transit times. Also of particular interest
   is the condition:
      \begin{equation} 
      \Rc = \Rc_{{\rm EV}} \equiv \frac{s_A}{s_B}
       \end{equation}
     which can be seen, from (A.2), to correspond to equal velocities $v_A = v_B$.
      Defining
    \begin{equation} 
  \Delta v \equiv v_A -v_B,~~~\bar{v} \equiv \frac{ v_A +v_B}{2}
      \end{equation}
   and combining (A.2), (A.4) and (A.5) gives
      \begin{equation} 
      \Delta v = \frac{2 \bar{v}(\Rc_{{\rm EV}} -1)}{(\Rc_{{\rm EV}}+\Rc)} =
      \frac{\bar{v}(\Rc_{{\rm EV}} -1)}{\Rc} +{{\rm O}}\left[\left(\frac{\Delta s}{\bar{s}}\right)^2\right]
       \end{equation}
     where
     \begin{equation} 
  \Delta s \equiv s_A -s_B,~~~\bar{s} \equiv \frac{s_A +s_B}{2}.
      \end{equation} 
    It follows from the relativistic relations:
        \begin{equation} 
          v =\frac{pc^2}{E} =\frac{pc^2}{\sqrt{m^2 c^4+p^2c^2}}
  \end{equation}
       that
       \begin{equation}
      \Delta{p} = \frac{E^3}{m^2 c^6} \Delta v. 
     \end{equation}
      Retaining only terms of order $\Delta s/\bar{s}$ in the phase difference formula (6.11)
 gives
      \begin{eqnarray}
    \Delta \phi & = & \phi_A-\phi_B = \frac{(mc)^2}{\hbar}\left[\frac{s_B}{p_B}-\frac{s_A}{p_A}\right]
     \nonumber \\
        & = & \frac{(mc)^2}{\hbar}\left[\frac{\Delta p}{\bar{p}^2}- \frac{\Delta s}{\bar{p}}\right] + 
       {{\rm O}}\left[\left(\frac{\Delta s}{\bar{s}}\right)^2\right]  \nonumber \\
      & = & \frac{(mc)^2}{\hbar}\left[-\frac{\Delta s}{\bar{p}} +\frac{\bar{s}}{\bar{p}}\left\{
        1 +\frac{\bar{p}^2}{(mc)^2}\right\} \frac{(\Rc_{{\rm EV}}-\Rc)}{\Rc}\right] +
    {{\rm O}}\left[\left(\frac{\Delta s}{\bar{s}}\right)^2\right] 
         \end{eqnarray}
       where, in the last line, (A.6), (A.8) and (A.9) have been used.
       \par Setting $\Rc = \Rc_{{\rm EV}}$ (equal velocities, and so different production times
        in the paths A and B) gives
    \begin{equation} 
     \Delta \phi  =  - \frac{(mc)^2}{\hbar}\frac{\Delta s}{\bar{p}}~~~~({\rm equal~velocities}).
      \end{equation}
    Setting $\Rc = 1$ so that $t_A = t_B$ and the paths have the same initial time, but are executed
     with different velocities, gives
        \begin{eqnarray}
    \Delta \phi & = &  \frac{(mc)^2}{\hbar}\left[-\frac{\Delta s}{\bar{p}}+ \bar{s}\left\{\frac{1}{\bar{p}}
      +\frac{\bar{p}}{(mc)^2}\right\} \frac{\Delta s}{s_B}\right]  \nonumber \\
 & = & \frac{(mc)^2}{\hbar}\left[- \frac{\Delta s}{\bar{p}}+\frac{\Delta s}{\bar{p}}
      +\frac{\bar{p}\Delta s}{(mc)^2}\right] +
    {{\rm O}}\left[\left(\frac{\Delta s}{\bar{s}}\right)^2\right]  \nonumber \\
     & = &  \frac{\bar{p}\Delta s}{\hbar}~~~~({\rm equal~times}) 
         \end{eqnarray}
   where, in the last line, only terms of order $\Delta s/\bar{s}$ have been retained. Equations (A.11) and (A.12)
    are Eqs.~(6.14) and (6.13), respectively, of the main text.
 \pagebreak     
   
\end{document}